\documentclass[twocolumn,showpacs,preprintnumbers,amsmath,amssymb,aps]{revtex4}

\usepackage[unicode]{hyperref}
\usepackage[caption=false]{subfig}
\usepackage{graphicx}
\usepackage{color} 
\usepackage{graphicx}

\graphicspath{ {./images/} }
\usepackage{dcolumn}
\usepackage{bm}
\usepackage[caption=false]{subfig}

\newcommand{\ovl}[1]{\overline{#1}}
\newcommand{\qq}[1]{#1}
\let\hat=\widehat

\newcommand{\dd}{\mbox{d}}

\begin{document}

\preprint{APS/123-QED}

\title{The effects of competition between random sequential nucleation 
of point-sized seeds and island growth by adsorption of finite-sized grains}

\author{A.~Khanam$^1$, J.A.D.~Wattis$^2$ and M.K.~Hassan$^1$}
\email[1]{ KHassan@du.ac.bd  } 
\email[2]{ Jonathan.Wattis@nottingham.ac.uk}
\date{\today}

\affiliation{
$1$ University of Dhaka, Department of Physics, 
Theoretical Physics Group, Dhaka 1000, Bangladesh. \\
$2$ School of Mathematical Sciences, University of Nottingham, 
University Park, Nottingham NG7 2RD, UK.}

\begin{abstract}
We study random sequential adsorption of particles from pool onto 
a one dimensional substrate following ballistic deposition rules, 
with separate nucleation and growth processes occurring simultaneously.  
Nucleation describes the formation of point-sized seeds, and 
after a seed is sown, it acts as an attractor and grows in size  
by the addition of grains of a fixed-sized. 
At each time step either an already-nucleated seed can increase in size, 
or a new seed may be  nucleated. We incorporate a parameter $m$, 
to describe the relative rates of growth to nucleation. 
We solve the model analytically to obtain gap size distribution function 
and a general expression for the jamming coverage as a function of $m$. 
We show that the jamming coverage $\theta(m)$ reaches its maximum 
value $\theta(m)=1$ in the limit $m\rightarrow \infty$ following a power-law 
$\theta(\infty) - \theta(m) \sim m^{-1/2}$. We also perform extensive 
Monte Carlo simulation and find excellent agreement between analytic 
and numerical results.
\end{abstract}
 

\maketitle

\section{Introduction \label{intro-sec}}

The kinetics of a monolayer growth by surface adsorption, 
deposition of particles onto solid substrates, has been studied 
extensively over the last many decades \cite{ref.evans, ref.tarjus, 
ref.schaaf, ref.bartelt_review, ref.book_privman, ref.schaaf_review}. 
One of the reasons for this is that it finds applications that covers 
many topic in physics, chemistry, biology and other branches of 
science and technology. There are numerous examples, such as  
the adsorption of colloids, bacteria, protein or latex particles on 
solid surfaces, macromolecules on biological membranes, growth 
of atomic islands of metals or semiconductors, etc., which are of 
interest in science and technology \cite{ref.has1, ref-svs, ref-svs2, 
ref.has2, ref.has3, ref.finegold, ref.evans_surface}. 
The development of a theoretical understanding of the kinetics 
of adsorption poses many fundamental challenges owing to its 
non-equilibrium nature which means that the well-developed 
formalism of equilibrium statistical physics cannot be applied. 
One of the litmus tests of the degree of its complexity is that one 
can hardly make any progress analytically in more than one dimension;  
most analytical work remains confined to one spatial dimension 
\cite{ref.quasi_1, ref.quasi_2, ref.quasi_3}.  The simplest case of 
random sequential adsorption (RSA) process was that considered 
by Alfred Renyi in 1958 \cite{ref.renyi, ref-gonz, ref-hemmer}, 
which is known as the `car parking' problem, in which one-dimensional 
`cars' of a unfiorm length  $\sigma$ are `parked' sequentially at 
random available positions in a car park of length $L\gg\sigma$.  
After a sufficiently long time, all remaining gaps are shorter than 
$\sigma$ and no further adsorption is possible, at which point the 
expected coverage is given by Renyi's constant, $\theta_\infty = 
0.7475979...$.  Typically RSA is considered as irreversible and 
particle-particle interactions are accounted for by not allowing 
overlaps. This minimalist model captures many generic features 
of deposition phenomena. 

Many variants of the RSA model have been devised suits various 
special conditions occurring in real systems: for example, 
to model protein adsorption onto DNA, the case of binding with 
overlaps \cite{ref.em1}, and reversible binding \cite{ref.em2} 
have been analysed. 
In 1992 Talbot and Ricci proposed a ballistic deposition (BD)
model to take into account of strong self-attraction effects. 
They assumed that \qq{
when the particle is deposited on the substrate, rather than 
being stationary, it moves towards closest already adsorbed 
particle, increasing the size of that island} 
\cite{ref.talbot,ref.ps}.  
The attempt is rejected only when it fails to reach the 
surface.  In 1995, Pagonabarraga {\it et al.} proposed yet 
another variant, in which particle can reach the surface 
along an inclined direction \cite{ref.pagonabarraga}. 
Later, the deposition of competitive binary mixture of 
finite sized particles was been considered \cite{
ref.hassan_schimidt, ref.hassan_kurths}. The RSA of mixture 
of particles of different uniform sizes has been studied 
\cite{ref.hassan_mixture, ref.lr1}.  
The RSA model in presence of precursor-layer diffusion and 
desorption has also been studied where accelerated RSA and 
growth-and-coalescence has been considered as a special case 
\cite{ref.filipe,ref.filipe2}.   
An extension to two dimensional RSA was simulated by Purves 
{\em et al.}\ \cite{ref.lr2}.   Recently, a variant of RSA to 
the adsorption of particles modelling mobile patches has been 
proposed motivated by the coverage of oil droplets by 
DNA-functionalized colloidal particles \cite{ref.mobole_patches}. 
The shapes of the depositing particles has also been considered 
in one of the recent articles \cite{ref.Adrian_Baule}. 
Besides, Subashiev {\it et al.} studied the RSA of shrinking or 
expanding particles \cite{ref.Subashiev}

In this article, we propose a model that generalises RSA to 
describe two processes: (i) the nucleation of point-sized 
seeds and (ii) the growth by the addition fixed-sized grains. 
Our aim remains that of understanding the kinetics through 
which a monolayer is formed. However, the distinction between 
the rates of nucleation and growth may lead to dynamics which 
differ significantly from Renyi's classic `car parking' problem.  
In particular, we assume that in each time interval either a 
seed is sown or an already-sown seed grows.  
We include a tuning parameter, $m$, which regulates the rate of 
growth of existing domains relative to the nucleation of new sites.  

We solve the model analytically to find an approximate expression 
for the coverage and for the jamming limit as a function of $m$. 
For large values of $m$ island growth dominates, through the 
frequent ballistic deposition of grains; whilst for small $m$, 
growth occurs at a relatively small rate, and the nucleation 
of fresh seeds is dominant. 
We also give an algorithm to simulate the model numerically. 
We find excellent agreement between our analytical and numerical 
results revealing that the proposed integro-differential equation 
of the model accurately describes the behaviour of the algorithm. 
We show that the jamming coverage reaches the maximum jamming 
coverage $\theta (m)\rightarrow 1$ in the limit $m \rightarrow \infty$ 
following a power-law $\theta(m)\sim 1- m^{-1/2}$.

The paper is organized as follows: in Sec. \ref{model-sec}, the model 
is introduced and the algorithm for simulating it is described. 
Sec. \ref{theory-sec}, contains a theoretical approach to solving 
the model to find the gap size distribution function and the behaviour 
in the jamming limit. Numerical results are presented in \ref{num-sec} 
alongside a comparison with the analytical solutions.  The results 
are discussed and conclusions drawn in Sec. \ref{conc-sec}.

\section{Definition of the model \label{model-sec}}

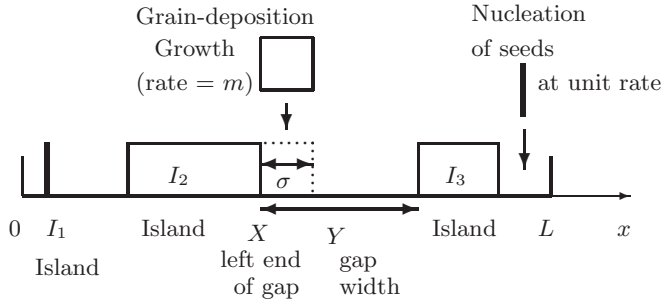
\begin{figure}[bt] 
\begin{picture}(200,120)(20,-33)
\put(0,0){\vector(1,0){230}}
\thicklines
\put(0,0){\line(0,1){15}}
\put(0,0){\line(1,0){200}}
\put(-5,-15){0}
\put(225,-15){$x$}
\put(200,0){\line(0,1){15}}
\put(195,-15){$L$}
\put(10,0){\line(0,1){20}}
\put(9,0){\line(0,1){20}}
\put(9,-15){$I_1$}
\put(5,-30){Island}
\put(40,0){\line(0,1){20}}
\put(40,20){\line(1,0){50}}
\put(90,0){\line(0,1){20}}
\put(45,-15){Island}
\put(55, 5){$I_2$}
\put(85,-17){$X$}
\put(75,-27){left end }
\put(80,-37){of gap}
\put(110,-5){\vector(-1,0){20}}
\put(110,-5){\vector( 1,0){40}}
\put(115,-18){$Y$}
\put(120 ,-27){gap}
\put(120 ,-37){width}
\put( 96,3){$\sigma$}
\put(100,12){\vector( 1,0){10}}
\put(100,12){\vector(-1,0){10}}
\multiput(110,20)(-3,0){7}{\circle*{1}}
\multiput(110,20)(0,-3){7}{\circle*{1}}
\put(100,35){\vector(0,-1){10}}
\put( 90,40){\line(1,0){20}}
\put( 90,40){\line(0,1){20}}
\put( 90,60){\line(1,0){20}}
\put(110,40){\line(0,1){20}}
\put( 43,65){Grain-deposition }
\put( 50,53){Growth}
\put( 43,40){(rate = $m$)}
\put(150,0){\line(0,1){20}}
\put(150,20){\line(1,0){30}}
\put(180,0){\line(0,1){20}}
\put(155,-15){Island}
\put(160,5){$I_3$}
\put(190,25){\vector(0,-1){15}}
\put(189,30){\line(0,1){20}}
\put(190,30){\line(0,1){20}}
\put(170,66){Nucleation }
\put(170,53){of seeds}
\put(195,40){at unit rate}
\end{picture}
\caption{Schematic description of the processes occurring in our 
model at the global scale; random sequential nucleation of 
seeds is shown on the right, and island growth deposition of a grain 
of positive size in the centre.  
\label{diagram-fig}}
\end{figure}

We introduce a new class of RSA which consists of two 
processes: firstly, {\em nucleation}---in which a point-sized 
{\em seed} is deposited in a gap on a substrate, and subsequently, 
{\em growth}---through which the island domains grow by the 
deposition of a {\em grain} of a positive size.  
Only one such event can occur in each time interval. 
The nucleation step is identical to classic RSA in that it 
occurs randomly, at any point in a gap, with equal probability. 
The deviation from RSA is that the seeds have zero size. 
Growth occurs through the deposition of domains of finite 
size $\sigma>0$, as in RSA, though these are {\em only} 
deposited ballistically - so that they are adjacent to an 
existing domain (whereas in RSA, deposition is typically 
at a randomly-determined location in a gap).  Any 
attempt to deposit a seed or a grain on an existing island 
is ignored. 

\begin{figure}[!hbt] 
\begin{picture}(400,200)(-20,-90)
\thicklines
\put(  0, 80){\line( 1,0){200}}
\put(  0, 90){\line( 1,0){60}}
\put( 60, 80){\line( 0,1){10}}
\put(140, 80){\line( 0,1){10}}
\put(200, 90){\line(-1,0){60}}
\put( 57, 60){$X$}
\put(130, 60){$X\!+\!Y$}
\put( 95, 65){$Y$}
\put(100, 75){\vector( 1,0){40}}
\put(100, 75){\vector(-1,0){40}}
\multiput(  0,80)( 5,0){12}{\line(1,2){5}}
\multiput(195,80)(-5,0){12}{\line(1,2){5}}
\put(  5, 95){Left Island}
\put(150, 95){Right Island}
\put( 75, 90){Gap -- size $Y$}
\put( 20, 45){the gap is treated as having a larger {\em effective}}
\put( 60, 32){size of $Y+2m\sigma$}
\put(  5, 27){Left }
\put(  5, 15){Island}
\put(175, 27){Right }
\put(175, 15){Island}
\put(  0, 0){\line( 1,0){200}}
\put(  0,-1){\line( 1,0){200}}
\put( 60, 0){\line( 0,1){15}}
\put(140, 0){\line( 0,1){15}}
\put(  0,10){\line( 1,0){60}}
\put(200,10){\line(-1,0){60}}
\multiput(  0, 0)( 5,0){12}{\line(1,2){5}}
\multiput(195, 0)(-5,0){12}{\line(1,2){5}}
\put( 85,-18){$Y+2m\sigma$}
\put(100, -7){\vector( 1,0){75}}
\put(100, -7){\vector(-1,0){75}}
\put( 25,-10){\line(0,1){25}}
\put(175,-10){\line(0,1){25}}
\put( 15,-20){$X\!-\!m\sigma$}
\put(160,-20){$X\!+\!Y\!+\!m\sigma$}
\put( 57, 20){$X$}
\put(130, 20){$X\!+\!Y$}
\put(100,-30){\vector( 1,0){39}}
\put(100,-30){\vector(-1,0){39}}
\put( 45,-30){\vector( 1,0){15}}
\put( 45,-30){\vector(-1,0){20}}
\put(160,-30){\vector( 1,0){15}}
\put(160,-30){\vector(-1,0){20}}
\put( 95,-42){$Y$}
\put( 38,-42){$m\sigma$}
\put(150,-42){$m\sigma$}
\put( 20,-55){Ballistic }
\put( 20,-66){Deposition }
\put( 20,-77){of grain }
\put( 20,-88){(Growth)}
\put(145,-55){Ballistic }
\put(145,-66){Deposition }
\put(145,-77){of grain }
\put(145,-88){(Growth)}
\put( 80,-55){Random}
\put( 80,-66){Adsorption}
\put( 80,-77){of seed}
\put( 80,-88){(Nucleation)}
\end{picture}
\caption{Illustration of the quantities involved 
	in one deposition event in our model.  
	\label{fig:2}}
\end{figure}
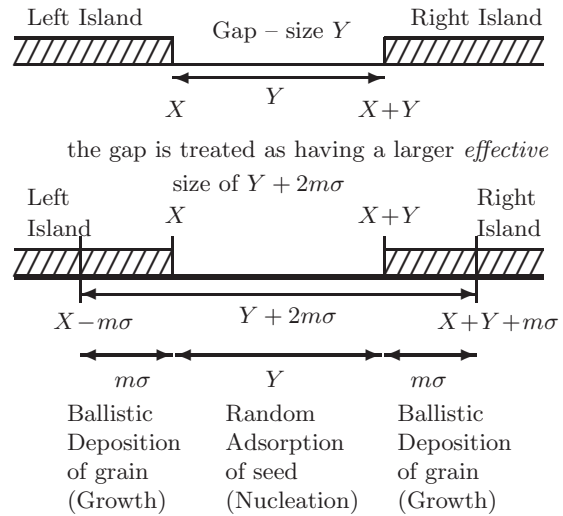

The islands thus have sizes which increase from zero (for a 
newly nucleated domain) in steps of $\sigma$ with the only 
upper bound being the size of the substrate simulated. 
The length of a gap is defined by the difference in 
locations of its ends, whether these are a seed, a large island, 
or the end of the domain. 

The only quantity we have yet to specify is the relative 
frequency of growth (deposition of $\sigma$-sized block) 
to nucleation (seed deposition). 
We denote this my $m$ so that when $m>1$ growth dominates 
and for $m<1$ nucleation occurs more frequently. 
These processes are illustrated in Fig.~\ref{diagram-fig}. 
The process may be simpler to understand by giving the algorithm
used in our simulations.  We start with a single empty substrate 
of length $L\gg1$, that is, a single gap. 
{\sl Commentary on the effect of steps is given in slanted text.}

\begin{itemize}
	
	\item[{\bf (i)}] Generate a random number, $R_1$, from a uniform 
	distribution on the interval $(0,L)$. 
	
	\item[{\bf (ii)}]  If $R_1$ falls in a gap, let $X$ be the location 
	of the left-hand edge of gap, and $Y$ be the length of the gap. 
	If $R_1$ falls on an existing island, go to step {\bf (iv)}. 
	
{\sl This method of choosing points for deposition 
means that larger gaps are chosen preferentially.} 

{\sl The role of $R_1$ is simply to choose a gap in which to deposit 
a particle, we next determine what is deposited in the gap and 
where it is deposited. } 		
	
\item[{\bf (iii)}] Determine whether nucleation or growth occurs:   \\ 
In place of the gap which actually occupies the interval $(X,X+Y)$, 
consider the gap as being the larger interval $(X-m\sigma,X+Y+m\sigma)$. 

Let $R_2$ be a random variable taken from a uniform 
distribution on the interval $(X-m\sigma,X+Y+m\sigma)$.  

{\sl A gap can be terminated by two seeds, two grains or a grain 
and a seed (or in the two cases $x=0$ and $x=L$, the end of the 
domain and either a seed or a gap):  how a gap is terminated 
has no influence on what deposition processes occur in the gap. }
	
\begin{description}

\item[{\bf (iii.a)}] 
	If $X-m\sigma < R_2 <X$ then deposit grain of size $\sigma$ 
	in the location $(X,X+\sigma)$; 
	{\sl the effect of this is to reduce the gap from the interval 
	$(X,X+Y)$ to $(X+\sigma,X+Y)$.  This constitutes growth the 
	left island by ballistic deposition of a grain of size $\sigma$.}

\item[{\bf (iii.b)}] 
	If $X<R_2<X+Y$  then deposit a seed nucleus (size zero) at $x=R_2$
	{\sl the effect of this is to replace the gap on the interval 
	$(X,X+Y)$ with two gaps, which are given by the intervals 
	$(X,R_2)$ and $(R_2,X+Y)$. This constitutes random sequential 
	nucleation of a zero-sized seed. }
	
\item[{\bf (iii.c)}] 
	If $X+Y<R_2<X+Y+m\sigma$ then deposit grain of size $\sigma$ 
	in the location $(X+Y-\sigma,X+Y)$; 
	{\sl the effect of this is to reduce the gap from the interval 
	$(X,X+Y)$ to $(X,X+Y-\sigma)$. This constitutes growth of the 
	right island by ballistic deposition of a grain of size $\sigma$.}

\end{description}
	
{\sl The effect of this is that probability of nucleation 
in this gap is $Y/(Y\!+\!2m\sigma)$ and the probability of 
growth in this gap is $2m\sigma/(Y\!+\!2m\sigma)$ 
so the relative rate of growth to nucleation is $2m\sigma/Y$.  
	
The effect of large values of $m$ is to make island growth by 
deposition of grains occur at a faster rate relative to nucleation 
of fresh seeds;  and, conversely, the effect of small $m$ is to 
make growth occur at a slower rate, thus favouring nucleation.  
The quantity $m$ can be viewed as a relative rate parameter. 
	
The reason we describing  grain deposition as ballistic, is 
that we assume the grain, once deposited on the substrate is 
mobile, and is attracted to the already-deposited grains, thus 
it moves to the closest endpoint (whether grain or seed) and 
stops moving when it comes into contact with a grain or a seed.} 
	
\item[{\bf (iv)}] Increase time by one and repeat from (i). 
{\sl It is important to increase time even when the attempt at 
deposition fails, since we are interested in the dynamics 
of the process, and how it slows at later times.}
				
\end{itemize}

\section{Analytical approach \label{theory-sec}}

\subsection{Derivation of mean field model}

To solve the model analytically, we need to appreciate the 
fact that the random sequential nucleation of point-sized 
seeds is equivalent to binary fragmentation of gaps.  
Placing a seed in a gap divides a gap of size $x$ into two 
smaller gaps, one of size $y$ and the other of size $x-y$ 
(with $0<y<x$).  In general, binary fragmentation is 
described by the equation  
\begin{eqnarray} 
\label{bin-frag-eq}
\frac{\partial c(x,t)}{\partial t} & = & 
-c(x,t)\int_0^x F(y,x-y) \, \dd y \\ \nonumber 
&& + 2 \int_x^\infty F(x,y-x)c(y,t) \, \dd y,
\end{eqnarray} 
where $c(x,t)$ denotes the expected number of gaps of 
length $x$ at time $t$.  
The first term on the right hand side of (\ref{bin-frag-eq}) 
describes the rate at which a particle (gap) of size $x$ is 
divided into two particles of size $y$ and $x-y$, and the 
second term describes a particle of size $y\geq x$ breaks 
into two smaller particles of sizes $x$ and $y-x$ \cite{
ref.krapivsky_redner_naim}. The choice of the kernel $F(x,y)$ 
describes the rate at which the particles fragment. The case 
$F(x,y)=1$ is known as Yule-Furry process \cite{ref.Bhabha}, 
or a random scission process \cite{ref.Jellinek}. The 
corresponding rate equation for the random sequential 
nucleation of point-sized seeds is therefore 
\begin{equation}
\frac{\partial c(x,t)}{\partial t} = 
-x c(x,t)+ 2 \int_x^\infty c(y,t) \, \dd y,
\end{equation}
where $c(x,t)$ is the gap size distribution function. 

To describe the creation of a gap of size $x$ due to growth 
of an island into a gap of size $x+\sigma$  we add 
a term of the form $2m\sigma c(x+\sigma,t)$.  Gaps of size 
$x$ are also destroyed due to growth of islands on both sides, 
so it is described by $2m\sigma c(x,t)$. Thus, the governing 
equation for $c(x,t)$ for random sequential nucleation and 
growth is given by the integro-differential equation
\begin{eqnarray}
\label{eq:1}
\frac{\partial c(x,t)}{\partial t} & =& 
-x c(x,t)+ 2 \int_x^\infty c(y,t) \,\dd y \\ \nonumber 
&& + 2m\sigma c(x+\sigma,t)-2m\sigma c(x,t), 
\\ && \nonumber \hspace*{3cm} 
\mbox{(for $x\geq \sigma$)}, \\
\frac{\partial c(x,t)}{\partial t} &=& 
-xc(x,t)+ 2 \int_x^\infty c(y,t)\, \dd y 
\nonumber \\ && 
+ 2m\sigma c(x+\sigma,t) , \qquad 
\mbox{(for $x<\sigma$)}. \label{eq:2}
\end{eqnarray} 
Note that each gap is either bordered by a grain or by a seed. 
The factor $m$ is added as tuning parameter
that tunes the strength of attraction for growth of grain.
This model describes the expected value of the stochastic 
algorithm given in Section \ref{model-sec}. Typically, 
there will be fluctuations, and the mean field results 
obtained by taking the average over many realisations 
of the algorithm.  

\subsection{Special case of the mean field model}

Combining (\ref{eq:1}) and (\ref{eq:2}), we obtain 
\begin{eqnarray}
\frac{\partial c}{\partial t} &=& - x c(x,t) 
+ 2 \int_x^\infty c(y,t)\,\dd y \\ \nonumber && 
+ 2 m \sigma c(x+\sigma,t) - 2 m \sigma c(x,t) H(x-\sigma) . 
\label{heav}
\end{eqnarray} 
In the limit $\sigma\rightarrow0^+$ we obtain 
\begin{equation}
\frac{\partial c}{\partial t}= -xc(x,t) + 2\int_x^\infty c(y,t)\,\dd y 
+ 2\mu \frac{\partial c}{\partial x},
\label{comb} 
\end{equation}
where $\mu = m \sigma^2$. 

The equation (\ref{comb}) can be solved by use of 
Laplace transforms, since 
\begin{eqnarray}
\ovl{c}(s,t) &=&  \int_0^\infty c(x,t) {\rm e}^{-sx} \,\dd x , \\ 
c(0,t) &=& \lim_{s\rightarrow\infty} s \ovl{c}(s,t) , 
\end{eqnarray}
imply that (\ref{comb}) 
can be rewritten as 
\begin{eqnarray}
\frac{\partial \ovl{c}}{\partial t} &=& 
\frac{\partial \ovl{c}}{\partial s} + 
\frac{2}{s} \left( \ovl{c}(0,t) - 
\ovl{c}(s,t) \right) + 2\mu s \ovl{c}(s,t) \nonumber \\ && - 
2\mu\lim_{\hat s\rightarrow+\infty} \hat s\,\ovl{c}(\hat s,t) . 
\label{Lap-eq}
\end{eqnarray}
If we assume $\ovl{c}(s,t)$ is a rational polynomial, namely 
$\ovl{c}(s,t) = A(t)/(s+B(t))$, then we find 
\begin{eqnarray}
\frac{\dd B}{\dd t} = 1 , &\quad &
\frac{1}{A} \frac{\dd A}{\dd t} = \frac{2}{B} - 2 \mu B , 
\end{eqnarray}
hence
\begin{eqnarray}
B(t) &=& t + \beta , \\ 
A(t) &=& \alpha (t+\beta)^2 {\rm e}^{-\mu t^2-2\mu\beta t},
\label{ABsol}
\end{eqnarray}
where $\alpha,\beta$ are arbitrary constants. 

In the case of $\sigma\rightarrow0^+$, the total coverage 
$\theta(t) = 1 - \int_0^\infty x c(x,t) \, \dd x$ is of course unity 
in the large time limit (that is, $\theta\rightarrow 1^-$ as 
$t\rightarrow+\infty$), but the kinetics by which this limit is 
approached is complicated, being given by 
\begin{eqnarray}
1-\theta(t) &=& \frac{1}{L} \int_0^L x \, c(x,t) \, \dd x 
\nonumber \\ &=& \frac{A(t)}{B(t)^2} = 
\alpha \,{\rm e}^{-\mu t^2 - 2 \mu \beta t} . 
\end{eqnarray}
Typically, this decay is extremely rapid (${\rm e}^{-\mu t^2}$);  
however, if $\mu$ were small then an intermediate timescale, 
where $1-\theta(t) \sim O(1/t)$, could be observed;  and if 
$\beta$ were large whilst $\mu$ small, then exponential decay 
might be seen ($1-\theta(t) \sim {\rm e}^{-\mu\beta t}$). 
In the case $\beta=0$ this simplifies a little, to $B = t$, 
$A = \alpha t {\rm e}^{-\mu t^2}$ and $1-\theta(t) = 
\alpha {\rm e}^{-\mu t^2}$. 

The form of the resulting solution of (\ref{comb})
\begin{equation}
c(x,t) = A(t) {\rm e}^{-xB(t)} , 
\label{eq:3}
\end{equation}  
motivates the use of this formula as an {\em ansatz} for the 
solution of the more general problem (\ref{eq:1})--(\ref{eq:2}) 
in which $\sigma>0$ and where we are interested in the 
solution at $\mathcal{O}(1)$ times.  
Dimensional analysis ($c_t \sim -xc$) implies that the gap 
size $x$ has the units of inverse time, thus $B(t)$ should 
have units of time so that the argument of the exponential 
function is dimensionless. Therefore we assume $B(t)\sim t$, 
taking $\beta=0$ since we do not expect the gap length 
distribution to be exponential at $t=0$. 

In order to obtain exact solutions for $A(t)$ we substitute 
the {\it ansatz} (\ref{eq:3}) into the rate equation (\ref{eq:1}) 
for $x>\sigma$, which (following application of $\dd B/\dd t=1$) 
gives the differential equation 
\begin{equation}
\label{eq:4}
\frac{\dd \ln A}{\dd t} = \frac{2}{t} - 2m \sigma (1-e^{-\sigma t}) , 
\end{equation}
for $A(t)$, whereupon $A(t)$ is given by  
\begin{equation} 
A(t) = A_0 t^2 \exp( - 2 m \sigma t+ 2m \sigma (1-e^{-\sigma t}) ). 
\label{Agensol}
\end{equation} 

\subsection{Solution of mean field model at early times}

The integro-differential equation (\ref{eq:1}) 
should be solved subject to the initial data 
\begin{equation}
c(x,0) = \delta(x-L) , \qquad  \mbox{with $L\gg1$} , 
\label{cic}\end{equation}
where $L$ is the length of the substrate. 
If we define moments of the distribution $c(x,t)$ by
\begin{equation}
I_n(t) = \int_0^\infty x^n c(x,t) \, \dd x , 
\end{equation}
then the initial conditions (\ref{cic}) imply
\begin{equation}
I_0(0) = 1 , \qquad I_1(0) = L . 
\label{IIC}
\end{equation}
However, given the {\em ansatz} (\ref{eq:3}) we have 
\begin{equation}
I_0(t) = \frac{A(t)}{B(t)} , \qquad I_1(t) = \frac{A(t)}{B(t)^2} , 
\end{equation}
thus for $t=0$ we have $B(0) =\beta$ and $A(0)=\beta^2$ 
giving $I_0(0)=\beta$ and $I_1(0)=\beta^2$.

This apparent contradiction between the initial conditions 
(\ref{cic}) and the {\em ansatz} (\ref{eq:3}) with (\ref{Agensol}) 
lies in considering an early time asymptotic analysis of 
the system.  At early times, since the substrate is almost 
entirely empty, there are a few gaps, gaps are typically 
large, and nucleation dominates growth, thus we have 
$I_0 = \mathcal{O}(1)$ and $I_1=\mathcal{O}(L)$. 
In this early time regime, we note that the gap shrinkage due 
to growth, $\sigma$, is infinitesimally small in comparison 
with the size of gaps, that is $\sigma \ll L$, and thus we 
approximate the dynamics of the system (\ref{heav}) by (\ref{comb}). 
By integrating, we note that, to leading order, we have 
\begin{equation}
\frac{\dd I_0}{\dd t} = I_1 - 2\mu c(0,t) , \qquad 
\frac{\dd I_1}{\dd t} = - 2 \mu I_0 , 
\end{equation}
together with (\ref{IIC}).  Thus, at early times, we have 
\begin{equation}
I_0 = \ovl{c}(0,t) \sim  1 + L t ,   \qquad 
I_1 \sim L - 2 \mu t . 
\end{equation}

Introducing the rescaling 
\begin{equation}
\tau = L t , \qquad q = s L , 
\label{qtau}
\end{equation}
with $\ovl{c}(p,0)={\rm e}^{-Lp}$ and $G(q,\tau) = \ovl{c}(p,t)$, 
we rewrite the governing equation (\ref{Lap-eq}) 
\begin{equation}
p \frac{\partial \ovl{c}}{\partial t} - 
p \frac{\partial \ovl{c}}{\partial p} = 
2 (\mu p^2 - 1) \ovl{c} + 2 I_0(t) - 2 p \mu c(0,t) , 
\label{Lap-eq2}
\end{equation}
at leading order, as
\begin{equation}
q \frac{\partial G}{\partial \tau} - 
q \frac{\partial G}{\partial q} = 2(1+\tau-G) , 
\label{Lap-eq3}
\end{equation}
with $G(q,0)={\rm e}^{-q}$. This can be solved by the 
method of characteristics, (giving ${\rm e}^u =1+\tau/q$ 
and $r=q+\tau$) which gives the solution   
\begin{eqnarray}
G(q,\tau) & = & (1 \!+\! \tau \!-\! q) + (q \!+\! \tau \!-\! 1 
+ {\rm e}^{-q-\tau} ) \left( \frac{q}{q\!+\!\tau} \right)^2 . 
\nonumber \\ & & \label{Gsol}
\end{eqnarray}

Inverting the transformations (\ref{qtau}), we find 
\begin{equation}
\ovl{c}(s,t) = \frac{Lt^2}{s+t} + \frac{t^2+2ts}{(t+s)^2}
+ \frac{ s^2 {\rm e}^{-Ls-Lt} }{(s+t)^2}, 
\end{equation}
which implies 
\begin{eqnarray}
\label{invtran}
c(x,t) & \sim & L t^2 {\rm e}^{-xt} +  (2-xt)t{\rm e}^{-xt} 
\\ \nonumber & & + H(x\!-\!L) [ {\rm e}^{-Lt} \delta(x\!-\!L) + 
t {\rm e}^{-xt} (xt\!-\!Lt\!-\!2) ] .  
\end{eqnarray}
Here, the leading order term, $c(x,t) \sim L t^2 {\rm e}^{-xt}$ 
implies $A(t) \sim L t^2$ for $t\ll 1$, so the constant $A_0$ 
in (\ref{Agensol}) should be given by $A_0=L$. 
Solving (\ref{eq:4}) for $A(t)$ subject to this 
matching condition gives
\begin{equation}
\label{eq:6}
A(t) = L t^2 e^{ - 2 m \sigma t + 2 m \sigma ( 1 - e^{-\sigma t} ) } .
\end{equation} 

The asymptotic solution of Eq.~(\ref{eq:1}) is therefore 
\begin{equation}
\label{eq:7}
c(x,t) =  L t^2 e^{2m(1-e^{-\sigma t})} {\rm e}^{-(x+2m\sigma)t}  
\hspace{3mm} {\rm for}  \hspace{3mm}  x \geq \sigma .
\end{equation}
The above solution implies that 
we can still recover the solution of the binary fragmentation 
process by taking $\sigma=0$ \cite{ref.krapivsky_redner_naim}.
It is important to mention that the solution for $c(x,t)$ alone is 
enough to provide us with all the interesting information that 
we need. 

\subsection{Behaviour of the coverage}

One of the most important quantities of interest in the 
process of RSA is the coverage, which is defined by 
\begin{equation}
\theta(t;m) = 1-\frac{1}{L} \int_0^L x\, c(x,t)\, \dd x .
\label{th-def}
\end{equation}

Denoting the units or dimensions of a term by $[\cdot]$, and 
considering the terms in Eq.~(\ref{heav}), we note that $[x]=1/[t]$, 
which is consistent with the term ${\rm e}^{xt}$ in (\ref{invtran}). 
Furthermore, since $L$ is length of the substrate, we have 
$[L]=[x]$, and $[\int x c(x,t) \, \dd x] =[x]$, thus $[c] = [x]^{-1} 
= [t]$, which is consistent with the prefactor on the right-hand 
side of Eq.~(\ref{eq:7}). 

Equation (\ref{th-def}) describes the fraction of substrate 
covered by deposited particles ($\theta$ being dimensionless). 
In the classic `car parking' case of RSA only 74.759 \% of the 
total substrate is covered.  We now investigate how the ballistic 
deposition of particles with random sequential nucleation changes 
the jammed state and kinetics of jamming. 

In the present case there is always space for nucleation 
as the seeds do not have any width (by definition). Seeds  
do not {\em directly} contribute to the coverage; however, 
they influence the jamming kinetics, since the presence of 
multiple nuclei within $\sigma$ of each other prevent the 
deposition of grains and reduce the rate of island growth. 

In this model the jamming coverage is the state when there 
are no more gaps available for the growth of particles of 
size $\sigma$.  Once this state is achieved, it is no longer 
necessary to continue considering the process of nucleation 
since the coverage can no longer change. 

\begin{figure}[!ht]  FIG 3 
\centering
\subfloat[]
{\includegraphics[height=5.5 cm, width=8.5cm, clip=true]
{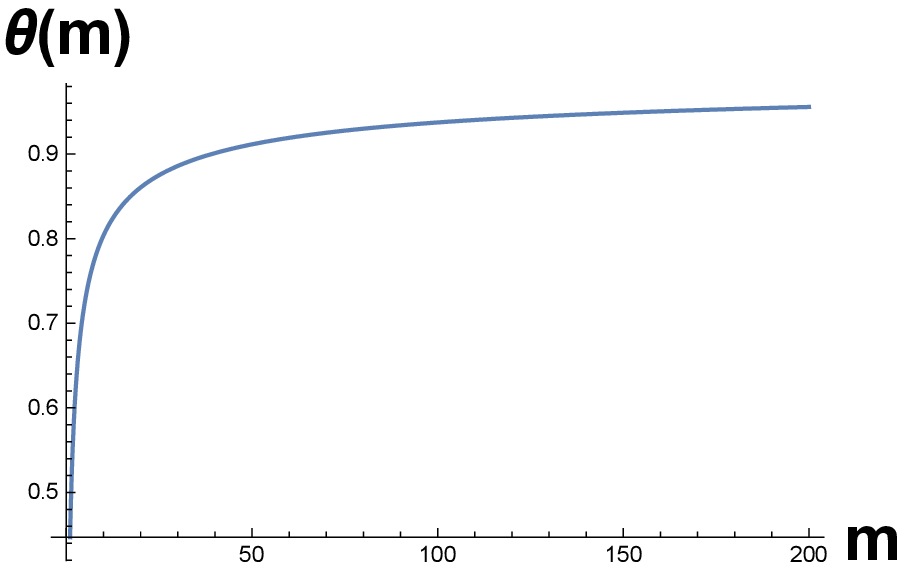} \label{fig:1a}}

\subfloat[]
{\includegraphics[height=5.5 cm, width=8.5cm, clip=true]
{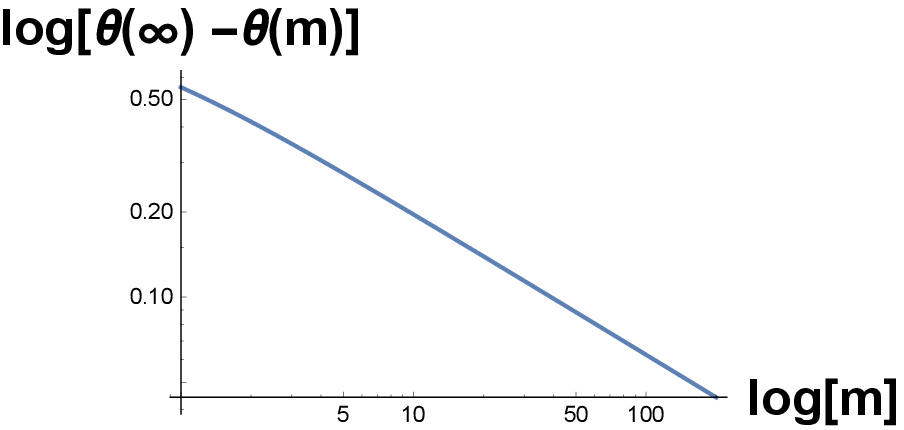} \label{fig:1b}}
\caption{(a) We show how $\theta(m)$ varies with $m$. 
(b) We plot $\log[\theta(\infty)-\theta(m)]$ versus $\log[m]$ 
and find perfectly straight line with slope almost equal 
to $1/2$ only if we choose $\theta(\infty)=1$, 
implying that the jamming coverage $\theta(m)$ 
reaches its maximum value of $\theta(m)=1$ following 
a power-law $\theta(m)\sim 1-m^{1/2}$. 
\label{fig:1ab}}
\end{figure}

It is interesting to determine how the kinetics of jamming 
differs in this model from that of standard RSA. 
To find the jamming coverage, it is more convenient 
to consider the rate equation for the coverage rather 
than the coverage itself. We differentiate the definition 
of $\theta$ with respect to time $t$, giving  
\begin{eqnarray}
\frac{\dd \theta}{\dd t} & = & - \frac{1}{L} \int_0^\infty 
x \frac{\partial c(x,t)}{\partial t} \,\dd x, 
\\ \nonumber &=& -\frac{1}{L} \int_0^\sigma x 
\frac{\partial c(x,t)}{\partial t} \,\dd x - \frac{1}{L}
\int_\sigma^\infty x \frac{\partial c(x,t)}{\partial t}\,\dd x.
\end{eqnarray} 
In the first term we substitute the rate equation for 
$x<\sigma$ and in the second term we substitute the 
rate equation for $x\geq \sigma$. Thus 
\begin{equation}
\frac{\dd\theta}{\dd t} = \frac{2 m \sigma^2}{L} 
\int_\sigma^\infty c(x,t) \,\dd x .
\label{dtheta}
\end{equation} 

This equation shows that only the finite-sized particles 
contribute to the coverage, and the lower limit of $\sigma$ 
since only gaps of size above $\sigma$ can be filled.  The 
right hand side of Eq.~(\ref{dtheta}) bears the dimension 
of inverse time, as it should. Since $\theta(t,m)|_{t=0}=0$, 
by (\ref{eq:3}) and (\ref{eq:6}) we find 
\begin{equation}
\theta(t, m)= 2m \int_0^{\sigma t} s \exp \left( 2m (1-e^{-s}) 
- (2m+1) s \right) \, \dd s,
\end{equation}
and hence
\begin{equation}
\label{eq:jamming_coverage}
\theta(\infty, m)= 2m \int_0^{\infty} s \exp \left( 2m (1-e^{-s}) 
-(2m+1) s \right) \,\dd s.
\end{equation}

Since $\theta$ is a dimensionless quantity and, in the limit 
of large substrate length $L$, the only length in the system 
is $\sigma$, thus the final coverage should be independent 
of $\sigma$  --a non-trivial and interesting result. In 
Fig.~\ref{fig:1a} we plot the analytical solution for jamming 
coverage given by (\ref{eq:jamming_coverage}) to show how the 
jamming coverage $\theta(m)$ varies with the strength of 
attraction $m$. 

To further quantify how the coverage $\theta$ depends on $m$, 
we plot $\log[\theta( t;m)|_{t\rightarrow\infty, m\rightarrow\infty} 
- \theta(t;m)_{t\rightarrow\infty} ]$ against $\log(m)$ 
in Fig.~\ref{fig:1b}, finding a straight line.  We only find 
a perfect straight line if we choose $\theta(\infty;m)|_{m 
\rightarrow \infty}=1$,  and then the slope is equal to $1/2$. 
This implies that $\theta(\infty;m)$ approaches $\theta(\infty,
m)_{m\rightarrow \infty}=1$ following the power-law 
$\theta(\infty, m)=1-m^{-1/2}$ for all $m$. 

\section{Numerical Simulation \label{num-sec}}

Here we address the questions such as `what are the dynamics 
of the process?', `what is the effect of the tuning parameter 
$m$?' Our goal is to verify and illustrate the theoretical 
results through numerical simulations of the nucleation and 
growth algorithm to show that the model is well understood. 

The simulations are restricted to a substrate of finite length, 
$L$. For sufficiently large $L$ (compared to the grain size, 
that is, $L\gg\sigma$) the effects of finite size $L$ are 
negligible.  We follow the algorithmic steps {\bf (i)}---{\bf (iv)} 
given in Section \ref{model-sec} which are also illustrated in 
Fig.~\ref{fig:2}.  The time scale in the simulations is given 
by the number of attempts to nucleate a seed or to grow a grain, 
regardless of whether the attempt is successful or not. 
Using this time scale and expressing all lengths in terms of 
$\sigma$, the simulation results can be directly compared to the 
solution of the rate equations (\ref{eq:1})--(\ref{eq:2}) and 
the solution for the jamming limit (\ref{eq:jamming_coverage}).

To test the analytic solution for $c(x,t)$ of the rate equation
(\ref{eq:1})--(\ref{eq:2}) given by (\ref{eq:7}), we need to 
choose a value for $m$ (eg $m=1$), perform a simulation until 
no further island growth is possible, and then extract the sizes 
of all gaps from the simulation at a range of times, $t$. 
The distribution of gap sizes can then be constructed. 

\begin{figure}[!ht] 
\centering
\subfloat[]{
\includegraphics[height=6.5 cm, width=8.5 cm, clip=true]
{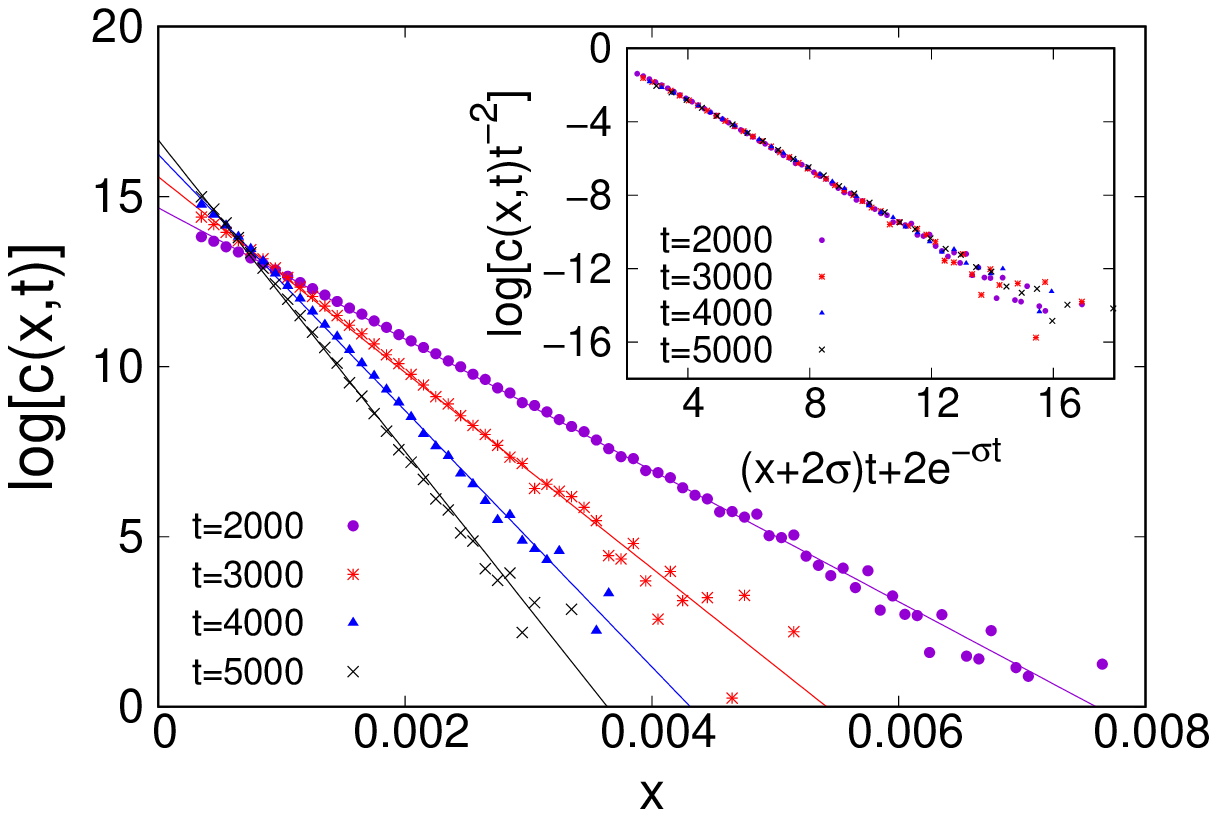}}
\caption{Plots of the gap size distribution function 
$c(x,t)$ against $x$ shown on a $\log$-linear scale
using ensemble averaged data for three different times.
The inset shows the same data, but with $\log[c(x,t)/Lt^2$ 
plotted against $(x+\sigma)t+2e^{-\sigma t}$ to show that 
the data collapses only a universal line.  
\label{fig:3}}
\end{figure}

This histogram data in which height represents the number of gaps 
within a given range of gap sizes (say of width $\Delta x$),   
normalized so that area under curve gives the number of gaps 
present in the system at time $t$ regardless of their sizes. 
Such data is shown in  Figs.~\ref{fig:3}, on a $\log$-linear scale 
so that the extremely small frequency of large gaps remains visible. 
Plotting the data at a range of times clearly shows a family of 
straight lines with time-dependent gradients and intercepts, 
indicating the distribution is exponential in $x$. 

\begin{figure}[!ht] 
\centering
\subfloat[]{
\includegraphics[height=6.5 cm, width=8.5 cm, clip=true]
{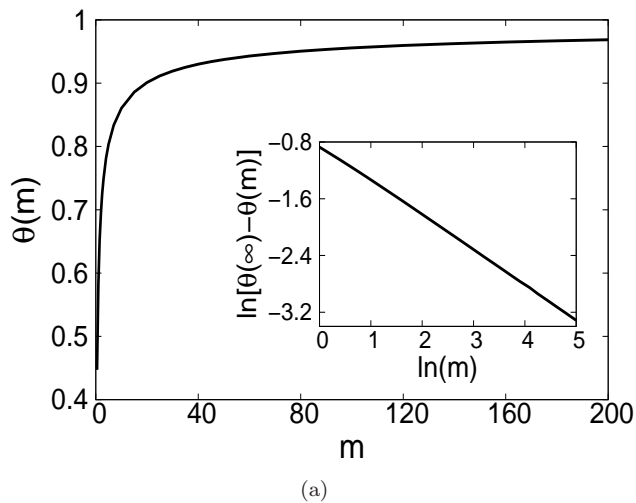}}
\caption{Plots of the jamming coverage $\theta(\infty; m)$ 
against $m$, clearly showing that as the strength of attraction
($m$) increases, the corresponding coverage increases sharply 
to its maximum value $\theta(\infty;m)_{m\rightarrow \infty}) 
\rightarrow 1$. 
\label{fig:4}}
\end{figure}

For further verification in the inset of Fig.~\ref{fig:3} we 
plot $\log[c(x,t)/Lt^2$ against $(x+2\sigma)t+2e^{-\sigma t}$ 
and find that all the data points at all times collapse onto 
a single straight line.   This is a strong confirmation of 
the theory of Section \ref{theory-sec}, and (\ref{eq:7}) in 
particular.   The inset of Fig.~\ref{fig:3} confirms that 
the universal distribution is exponential. 

The coverage in the jamming limit is one of the most interesting 
properties of generalised RSA processes. We find that the jamming 
coverage increases with the relative growth rate, $m$, as expected; 
this is shown in Fig.~\ref{fig:4}. In the limit $m\rightarrow \infty$, 
the jamming coverage approaches the maximum value of $\theta=1$. 
Comparing Figs.~\ref{fig:4} and \ref{fig:1ab}, we find an excellent 
agreement between simulation and theory. 
To determine {\em how} the jamming coverage $\theta(\infty,m)$ 
depends on $m$, we plot $\log(\theta(\infty;m)|_{m\rightarrow\infty}-
\theta(\infty;m)$ against $\log (m)$ in the inset of Fig.~\ref{fig:4} 
and find a straight line.  It is noteworthy that the final coverage 
is described by a formula as simple as $\theta=1-m^{-1/2}$. 

\section{Discussion and Conclusions \label{conc-sec}}

In this article, we have studied formation of monolayer through 
the nucleation of point-sized particles and the subsequent growth 
by fixed-sized grains. We have assumed that deposited domains, or 
'islands' act as attractors for grains (via ballistic deposition) 
and so increase in size, simultaneously with the nucleation of 
fresh sites by deposition at random vacant locations. As well as 
the size of the grains through which growth occurs ($\sigma$), we 
have also incorporated a parameter $m$ to control the strength of 
the attraction by increasing the apparent size of the seed or grain. 

From the algorithm we have derived an integro-differential equation 
which predicts the mean behaviour of the stochastic process. 
We have solved this model analytically determining an {\em ansatz} 
for the gap size distribution function (\ref{eq:3}) and hence 
an expression for the jamming coverage $\theta(m)$. 
Remarkably, an explicit asymptotic approximation (\ref{invtran}),  
for the gap size distribution in the early timescale can be found 
using Laplace transforms.  This shows how the delta function 
initial conditions (\ref{cic}) evolve to the intermediate dynamics 
and motivate the form of the {\em ansatz} (\ref{eq:3}) which  
explains the behaviour of the system in the approach to jamming.  
We have performed extensive Monte Carlo simulations based on an 
algorithm for the process and found excellent agreement between 
the numerical simulations and our analytical results. 

To summarize, we have generalised RSA to include separate 
nucleation and growth processes, incorporating a parameter, $m$, 
which defines relative rates of grain-growth to seed nucleation. 
We have solved the model analytically to find the gap size 
distribution function.   We have shown that the jamming coverage 
increases to complete saturation with $m$ following a power-law. 
Our results provide insights into the final coverage and how the 
morphology of the jamming state depends on $m$.



\end{document}